\documentclass[12pt]{article}
\usepackage{amssymb}
\usepackage{amsfonts}
\usepackage{amsmath,amsthm}
\usepackage{graphics,epsfig}
\usepackage{subfigure}
\usepackage[symbol]{footmisc}
\usepackage{tikz}
\usetikzlibrary{decorations.pathmorphing}
\usetikzlibrary{shadows,shapes,arrows,chains,decorations.pathreplacing,calc}

\oddsidemargin 10mm \numberwithin{equation}{section}
\def\be{\begin{equation}}
\def\ee{\end{equation}}
\begin{document}
\begin{center} {{\bf {Effects of quintessence dark energy on the action growth and butterfly velocity}}\\
 \vskip 0.50 cm
  {{ Hossein Ghaffarnejad \footnote{E-mail: hghafarnejad@semnan.ac.ir
 } }{ Mohammad Farsam \footnote{E-mail: mhdfarsam@semnan.ac.ir
 } }{ Emad Yaraie \footnote{E-mail: eyaraie@semnan.ac.ir
 } }}\vskip 0.2 cm \textit{Faculty of Physics, Semnan
University, P.C. 35131-19111, Semnan, Iran }}
\end{center}
\begin{abstract}
In this work we are about to investigate the effects of
quintessence dark energy  on evolution of the computational
complexity relating to the AdS/CFT correspondence. We use
$``complexity=action"$ conjecture for a charged AdS black hole
 surrounded by the dark energy  at the quintessence regime.
 Then we try to find some conditions on the quintessence parameters
 where the Lloyd bound is
 satisfied in presence of affects of the quintessence dark energy
 on the complexity growth at the late time approximations. We compare late time approximation of the action growth
 by perturbed geometry in small limits of shift function.
  Actually
 we investigate the evolution of complexity when thermofield double state on the boundaries is perturbed by local operator
 corresponding to a shock wave geometry as holographically.  Furthermore we seek spread of local shock wave on the black
  hole horizon in presence of the quintessence dark energy.
\end{abstract}

\section{Introduction}
In the perspective of gauge/gravity duality evolution of all
dynamical fields in the anti-de Sitter (AdS) gravity have dual
pictures in the boundary field theories on which the gravity has
been removed [1]. Actually this duality acts as a dictionary for
all field theory characteristics in the language of black hole
physics in AdS spacetime. One of important conjectures in the
holographic context is about computational quantum complexity
which implies the minimum quantum gates necessary to produce
states associated with boundary complexity from the reference
state. These conjectures are based on the behavior of a patch
created by the light rays emitted from $t_L$ on the left boundary
and $t_R$ on the right side of a two-sided eternal black hole,
called Wheeler-DeWitt (WDW) patch [2,3]. The old conjecture states
``complexity=volume" (CV), in which the volume of a maximal
space-like slice in the black hole interior that connects $t_L$
and $t_R$ supposed to be appropriated with
 computational complexity in its conformal field theory dual on the boundary [4]. This conjecture is a result of the behavior of the interior
 volume of black hole which grows linearly with time, so it could be translated with the growth of computational complexity on the boundary with
  time [5,6]. However, if the bulk contains a shock wave the interior volume of the black hole shrinks for a finite time
  interval and shows an opposite
  behavior. In the new conjecture of ``complexity=action" (CA), computational complexity of a holographic state on
  the boundary pictures is  as
  the on-shell action in WDW patch.
  This new conjecture has some preferences with respect to the old one and solve some problems which the ``CV"
  conjecture suffers.
  Lloyd showed [7] the growth rate of quantum complexity has an upper bound which is related to the average
   energy of the orthogonal quantum states $E$ such that
\begin{equation}
\frac{d (Action)}{dt}\leq 2E.
\end{equation}
In this new conjecture, authors of the works [2,3] concluded that
the action growth of WDW patch obeys this bound at late time
approximation which is
 provided us to work with orthogonal states. At this approximation we can have a general and universal
 form of the above bound for a rotating charged black hole as follows
 [8].
\begin{equation}
\frac{d\mathcal{A}}{dt}\leq(M-\Omega J-\mu Q)_+-(M-\Omega J-\mu Q)_-,
\end{equation}
where $+$ and $-$ indicate the lowest and highest energy of
states. These states for a black hole with multiple horizon happen
at the most outer $(r_+)$ and the most inner $(r_-)$ horizons.

In the other side, shock waves near the horizon of an AdS black
hole describe chaos in a thermal CFT [9,10,11] and could have
interesting effects on the boundary complexity. From the point of
view of holography dictionary the spreading of the shock wave near
the horizon has butterfly effect in the boundary field theory.
This effect could be seen with a sudden decay after scrambling
time
   $t_*=\frac{\beta}{2\pi}{ln}S$, in which
  ``$S$" stands for the black hole entropy.
 It is the necessary time for the black hole as the fastest scramblers to render the density matrix of a small
  essentially exactly thermal subsystem.
   The spreading local
    shock waves in the bulk arise from throwing a few quanta into the horizon which corresponds to perturb thermofield double state $|TFD>$ on the
     boundaries by local operators. The growth of spreading the shock wave on the boundary is identified with butterfly velocity
      ``$v_B$" could be
     obtained by solving the equation of motions of perturbed geometry.
Complexity growth rate and the effects of butterfly on it, are
investigated on various gravity models for the bulk. In refs.
[2,3,8] the authors investigated action growth for various AdS
black holes and tested the Lloyd bound by considering the effects
of charge. The growth of holographic
 complexity is studied in massive gravity in ref. [12], and in a more variety of other works [13-19].
 On the other side, some of works have been done about   studying  the shock wave geometry in different
  gravity models in the bulk [9,10,11,20,21] and  effects of them were investigated on the action growth by obtaining
   butterfly velocity and comparing them with other simple models  [22,23]. \\
  Motivation of studying the effect of dark energy arises from several works in holographic context.
  For instance Chen et al  found that
   quintessence dark energy can affect  the s-wave and p-wave holographic superconductor  [24]. In the other side, Kuang
    et al. studied the
   holographic fermionic spectrum dual to AdS black brane in 4 dimensions in the presence of quintessence dark energy and showed that
   this fermionic system exhibits a non-Fermi liquid behavior [25].  So it would be natural to investigate other aspects of holographic
   effects of this quintessence dark energy such as its impact on complexity growth or its effect on the spread of chaos on boundary.
   Quintessence dark energy introduced by an equation of state arisen from its energy tensor has a
state parameter which is varied like $-1<\omega<-\frac{1}{3}$.
This state parameter which is a factor to explain the accelerating
expansion of the universe, could define various regimes and could
be fixed by regarding some cosmological observations. Here we like
to study effects of this factor on the holographic complexity of a
AdS black hole which is perturbed with a shock wave matter field.
 It can help us to get more profound understanding from the entropy of the black hole in presence of the quintessence dark
 energy by attention to some related bounds like Lloyd bound.
  Also it could help us to have more information about quintessence state
parameter to get a better understanding of late time acceleration
of the universe. Furthermore it could be interesting how the
spreading of a chaos could be sensitive for changes of $\omega$.
This could give a comprehensive and statistical study for
different regimes of the used gravity theory during the evolution
of the  action. Moreover, from [26] we know free parameter $a$
emerged from quintessence energy tensor should be considered as
thermodynamic variable and so its physical insight needs to
investigation more. So as next work we are interested to study
dual CFT perspective of this variable to approach to goal of this
article.

  In the present work we consider the effects of quintessence dark energy on the AdS black holes geometry, and therefore we will see how it changes the
action growth rate and butterfly velocity in the shock wave
geometry. Quintessence dark energy is a canonical scalar field
which is one of the successful theories to explain the
acceleration phase of the universe [27,28,29]. In this model which
was first introduced by Kiselev [30] an additional energy-momentum
 tensor of quintessence counterpart must be added to the Einstein equation as $\mathcal{G}=\kappa(T_{matter}+T_{quintessence})=\kappa\mathcal{T}$.
  The effects of quintessence
   have been studied in a wide range of works and thermodynamics of the various black holes
   have been investigated when they are surrounded by the dark energy
  [30,31,32,33,34,35]. It would be challenging to see how it affects the holographic characteristics as
  well. Layout of this work is as follows.\\
We first study the action growth in the presence of quintessence
dark energy in section 2. Then we discuss about conditions where
the Lloyd  bound [7] could be hold with new charge associated with
this new field. In section 3 we calculate the butterfly velocity
related to the spreading  of perturbation and compare the action
growth in the presence of a local shock wave geometry in the
gravity model under consideration.
 Section 4 denotes to concluding remarks and outlook of the work.

\section{The rate of action growth in presence of the dark energy}
We consider RN-AdS black hole surrounded by the quintessence dark
energy in four dimensional curved space time. It could be
described by the following action functional.
\begin{equation}
\mathcal{S}=\mathcal{S}_{bk}+\mathcal{S}_{bd}
\end{equation}
where the first part is related with the bulk action contains
Einstein-Maxwell Lagrangian density defined in the AdS spacetime
as follows.
\begin{equation}
\mathcal{S}_{bk}=\frac{1}{16\pi G}\int_{bulk}
d^4x\bigg(\sqrt{-g}\big[R-2\Lambda-F^{\mu\nu}F_{\mu\nu}\big]
+\mathcal{L}_q\bigg).
\end{equation}
In the above action the cosmological constant is related to the
AdS space radius $L$ by $\Lambda=-3/L^2$ in four dimension. The
second term in the action (2.2), $\mathcal{L}_q$ implies on the
lagrangian of the quintessence dark energy as a barotropic perfect
fluid defined by [36]
\begin{equation}
\mathcal{L}_q=-\rho c^2\bigg(1+\omega\ln\big(\frac{\rho}{\rho_0}\big)\bigg),
\end{equation}
in which $c$ is the light speed and $\rho_0$ is integral constant
which is come from singularity cut-off. Also the quintessence dark
energy barotropic index satisfies $-1<\omega<-\frac{1}{3}$. It
comes from the quintessence dark energy equation of state as
$p=\omega\rho c^2$ in which $\rho$ is energy density and $p$ is
corresponding isotropic pressure.

In the other side the Gibbons-Hawking-York (GHY) boundary part of
the action term within the WDW patch at the late time
approximation is defined by
\begin{equation}
\mathcal{S}_{bd}=\frac{1}{8\pi G}\int_{boundary}d^3x\sqrt{{-h}}K,
\end{equation}
where $h$ stands for the determinant of induced metric on the
boundary of AdS bulk and $K$ represents the trace of extrinsic
 curvature. We should note the authors of the work [37] showed
 that ``CA" conjecture suffers from some ambiguities related
 to the null surface's parametrization and they found joint and boundary terms which are absent in this proposal.
 However authors of the works [38,39]
 proved that the action growth at late time approximation does not need these extra terms.
 So for the present work, all other boundary
 terms and joint terms vanish at late time approximation.

 In general, a line element for spherically symmetric static geometry
 in the Schwarzschild coordinates could be given by
\begin{equation}
ds^2=-f(r)dt^2+\frac{dr^2}{f(r)}+r^2{d\vec{x}_2}^2,
\end{equation}
with the following solution in presence of the quintessence dark
energy effect [30].
\begin{equation}
f(r)=1-\frac{2M}{r}-\frac{\Lambda r^2}{3}+\frac{q_E^2}{r^2}-\frac{a}{r^{3\omega+1}},
\end{equation}
where the positive constant $``a"$ treats as normalization factor
for density of the quintessence dark energy via
$\rho=-3a\omega/2r^{3(\omega+1)}$. The electromagnetic tensor
field  $F_{\mu\nu}$ is defined by 4-vector potential $A_{\mu}$ as
$F_{\mu\nu}=\partial_\mu A_{\nu}-\partial_\nu A_\mu.$ This is a
gauge field and so for a spherically symmetric static metric (2.5)
we can take  its form as spherically symmetric time independent
function for simplicity reasons as follows.
\begin{equation}
A=A_tdt=-q_E(\frac{1}{r}-\frac{1}{r_+})dt.
\end{equation}
Substituting the metric solution (2.5) with (2.6) one can
calculate the Ricci scalar as
\begin{equation}
R=4\Lambda+\frac{3a\omega(3\omega-1)}{r^{3(\omega+1)}}.
\end{equation}
To study the evolution of WDW action of this model it is useful to
depict its Penrose diagram at late
 time approximation. As the  black hole solution (2.5) has multiple horizons so we must consider its behavior between the lowest and the
 highest energy which happen at $r_-$ and $r_+$, respectively.
 In figure 1, we depicted the evolution of WDW patch for black hole containing the multiple
horizons at late time approximation
 [40].
  By increasing the time on the boundary the patch terminates at location of $r=r_{meet}(t_L,t_R)$ for
  all charged black holes.
 Time transition says us that the action growth at late time
 approximation only relates to the dark blue region behind the future horizon in figure 1.
   Of course, the tiny part above the meet line is in second order $\delta^2$ which is negligible.

\begin{figure}[h]
\begin{center}
\begin{tabular}{cc}
\setlength{\unitlength}{1cm}

\begin{tikzpicture}[scale=1.2]

\draw [thick]  (0,0) -- (0,3); \draw
[thick,decorate,decoration={zigzag,segment
length=1.5mm,amplitude=.3mm}]  (0,3) -- (0,5); \draw [thick] (0,0)
-- (0,-1.5); \draw [thick]  (3,0) -- (3,3); \draw
[thick,decorate,decoration={zigzag,segment
length=1.5mm,amplitude=.3mm}]  (3,3) -- (3,5); \draw [thick] (3,0)
-- (3,-1.5);

\draw [thick,dotted]  (3,3) -- (.95,5); \draw [thick,dotted]
(0,3) -- (2.05,5); \draw [thick,dotted]  (0,0) -- (1.5,-1.5);
\draw [thick,dotted]  (3,0) -- (1.5,-1.5); \draw [thick,dotted]
(0,0) -- (-1.5,1.5); \draw [thick,dotted]  (-1.5,1.5) -- (0,3);
\draw [thick,dotted]  (3,0) -- (4.5,1.5); \draw [thick,dotted]
(4.5,1.5) -- (3,3);
\draw [thick,dotted]  (-1.5,4.5) -- (-1,5); \draw [thick,dotted]
(3,3) -- (4.5,4.5); \draw [thick,dotted]  (4.5,4.5) -- (4,5);

\draw [thick,dotted]  (-1.5,-1.5) -- (3,3); \draw [thick,dotted]
(0,0) -- (-1.5,1.5); \draw [thick,dotted]  (-1.5,1.5) -- (0,3);

\draw [thick,dotted]  (4.5,-1.5) -- (-1.5,4.5);

\draw [thick,blue]  (1.5,3.2) -- (3,1.7); \draw [thick,blue]
(3,1.7) -- (1.5,0.2);

\clip (3,0.2) rectangle (0,6.2); \fill[fill=blue!50,opacity=.5]
(0.1,1.8) node {} -- (1.5,3.2) node {} -- (1.4,3.3) node {} --
(0,1.9) node {};

\draw [thin,dashed]  (0,3) to[out=15,in=165] (3,3);

\draw [thick,blue]  (0,1.9) -- (1.4,3.3); \draw [thick,blue]
(1.4,3.3) -- (1.5,3.2);

\clip (3,0.2) rectangle (0,3.2); \fill[fill=blue!10,opacity=.5]
(1.6,.3) node {} -- (3,1.7) node {} -- (1.5,3.2) node {} --
(0.1,1.8) node {};

\clip (3,0.2) rectangle (0,3.2); \fill[fill=red!20,opacity=.5] (1.5,0.2)
node {} -- (1.6,.3) node {} -- (.1,1.8) node {} -- (0,1.7) node
{};

\draw [thick,red]  (1.5,0.2) -- (0,1.7); \draw [thick,red]  (1.5,0.2)
-- (1.6,0.3); \draw [thick,red]  (0,1.7) -- (0,1.7); \draw
[thick,red]  (0,1.7) -- (1.5,3.2);

\draw [thick,red]  (1.5,3.2) -- (3,1.7); \draw [thick,red]  (3,1.7)
-- (1.5,0.2);

\draw [thick,blue]  (1.6,0.3) -- (0,1.9);

\draw [thick,violet]  (1.6,.3) -- (3,1.7); \draw [thick,violet]
(3,1.7) -- (1.5,3.2);

\draw [thick,dotted]  (3,0) -- (0,3); \draw [thick,dotted]  (0,0)
-- (3,3);

\end{tikzpicture}
\qquad\qquad & \hspace{1.5cm} \qquad
  \put(-214,106){\scriptsize $t_{L}\rightarrow$ }
  \put(-228,114){\scriptsize $t_{L}+\epsilon\rightarrow$ }
  \put(-90,107){\scriptsize $\leftarrow t_{R}$ }
  \put(-163,94){\rotatebox{45}{$r_2$}}
  \put(-135,98){\rotatebox{-45}{$r_2$}}
  \put(-163,127){\rotatebox{-45}{$r_2$}}
  \put(-138,117){\rotatebox{45}{$r_2$}}
  \put(-170,168){\rotatebox{45}{$r_{1}$}}
  \put(-122,170){\rotatebox{-45}{$r_1$}}
  \put(-180,22){\rotatebox{-45}{$r_0$}}
  \put(-114,19){\rotatebox{45}{$r_0$}}
  \put(-230,180){\rotatebox{-45}{$r_\infty$}}
  \put(-60,175){\rotatebox{45}{$r_\infty$}}
  \put(-220,15){\rotatebox{45}{$r_4$}}
  \put(-73,20){\rotatebox{-45}{$r_4$}}
  \put(-230,127){\rotatebox{45}{$r_3$}}
  \put(-225,70){\rotatebox{-45}{$r_3$}}
  \put(-61,130){\rotatebox{-45}{$r_3$}}
  \put(-65,68){\rotatebox{45}{$r_3$}}

\end{tabular}
\end{center}
\caption{ Penrose diagram for a two sided black hole with
multiple horizons in which WDW patch evolves at
 late time approximation and terminates at meet point with $t_L=t_R$ and $t=0$. $r_0$ is the null spatial infinity and the wavy
  lines indicate the singularities at $r=0$, also $r_\infty$ stands for $r=-\infty$. }
\label{fig-WDW}
\end{figure}
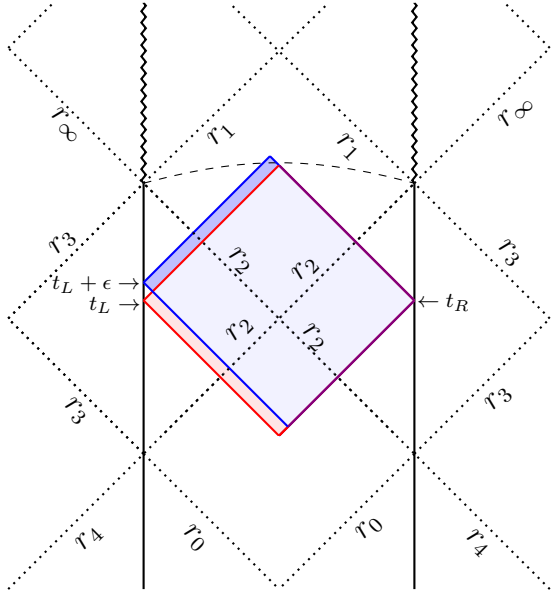

Now by studying time dependence behavior of total action (2.1) in
the quintessence regime of dark energy, we can evaluate
holographic complexity growth. By considering the Ricci scalar
(2.8) and the quintessence lagrangian density (2.3) one can
calculate the action growth of the bulk such that
\begin{equation}
\frac{d\mathcal{S}_{bk}}{dt}=\frac{1}{16\pi
G}\int\int_{r_-}^{r_+}r^2\bigg(2\Lambda+\frac{3a\omega(3\omega-1)}{r^{3(\omega+1)}}+\frac{2q_E^2}{r^4}$$$$
+\lambda(\omega)\frac{1}{r^{3\omega+1}}+\mu(\omega)\frac{\ln(r)}{r^{3\omega+1}}\bigg)dr{d\Omega_2},
\end{equation}
in which
\begin{equation}
\lambda(\omega)=-\frac{3a\omega}{2}(1+\omega\ln(-\omega)),~~~~
\mu(\omega)=\frac{9a\omega^2(\omega+1)}{2},
\end{equation}
and we put $\rho_0\equiv3a/2$. It is simple to show
$\lambda(\omega)>0$ and $\mu(\omega)>0$ in quintessence regime
$-1<\omega<-1/3$ as follows.  By setting $\Omega_2/4\pi G=1$ (2.9)
reads

\begin{equation}
\frac{d\mathcal{S}_{bk}}{dt}=-\frac{1}{2L^2}(r_+^3-r_-^3)-\frac{q_E^2}{2}\big(\frac{1}{r_+}
-\frac{1}{r_-}\big)-\frac{a(3\omega-1)}{4}\big(\frac{1}{r_+^{3\omega}}-\frac{1}{r_-^{3\omega}}\big)$$$$
-\frac{1}{12\omega}\bigg(\lambda+\frac{\mu}{3\omega}\bigg)\bigg(\frac{1}{r_+^{3\omega}}-\frac{1}{r_-^{3\omega}}\bigg)-
\frac{\mu}{12\omega}\bigg(\frac{\ln(r_+)}{r_+^{3\omega}}-\frac{\ln(r_-)}{r_-^{3\omega}}\bigg).
\end{equation}
It is also easy to see $\big(\lambda+\frac{\mu}{3\omega}\big)>0$
for quintessence regime $-1<\omega<-\frac{1}{3}$.

To obtain the action growth of the boundary part we must evaluate
the extrinsic curvature associated to the metric solution (2.5)
for which we have
\begin{equation}
K=\frac{1}{r^2}\frac{\partial}{\partial r}(r^2\sqrt{f(r)})=\frac{2}{r}\sqrt{f(r)}+\frac{f^{\prime}(r)}{2\sqrt{f(r)}},
\end{equation}
where the prime $\prime$ denotes to derivative with respect to
``$r$". By this definition time derivative of the second part of
the action (2.1) leads to  the following form.
\begin{gather}
  \frac{d\mathcal{S}_{bd}}{dt}=\frac{1}{8\pi G}\int_{boundary}{d\Omega_2}(\sqrt{-h}K) =\bigg[rf(r)+\frac{r^2f^{\prime}(r)}{4}\bigg]_{r_-}^{r_+} \\
  =(r_+-r_-)+\frac{3}{2L^2}(r_+^3-r_-^3)+\frac{q_E^2}{2}\big(\frac{1}{r_+}-\frac{1}{r_-}\big)+\frac{3a(\omega-1)}{4}\big(\frac{1}{r_+^{3\omega}}
  -\frac{1}{r_-^{3\omega}}\big).
\end{gather}
Adding (2.11) and (2.13) we obtain total growth action for
quintessence RN-AdS black hole such as follows.
\begin{equation}
\frac{d\mathcal{S}}{dt}=(r_+-r_-)+\frac{r_+^3-r_-^3}{L^2}-\frac{a}{2}\big(\frac{1}{r_+^{3\omega}}-\frac{1}{r_-^{3\omega}}\big)$$$$
-\frac{1}{12\omega}\bigg(\lambda+\frac{\mu}{3\omega}\bigg)\bigg(\frac{1}{r_+^{3\omega}}-\frac{1}{r_-^{3\omega}}\bigg)-
\frac{\mu}{12\omega}\bigg(\frac{\ln(r_+)}{r_+^{3\omega}}-\frac{\ln(r_-)}{r_-^{3\omega}}\bigg).
\end{equation}
Solving the  horizon equation $f(r_\pm)=0$ one can obtain the
following expressions for charge and mass of the RN-AdS black
hole.
\begin{equation}
q_E^2=r_+r_-\bigg[1+\frac{1}{L^2}\frac{r_+^3-r_-^3}{r_+-r_-}-a\frac{r_+^{-3\omega}-r_-^{-3\omega}}{r_+-r_-}\bigg],
\end{equation}
\begin{equation}
M=\frac{1}{2}\Bigg[(r_++r_-)+\frac{1}{L^2}\frac{r_+^4-r_-^4}{r_+-r_-}+\frac{a}{2}\frac{r_+}{r_-^{3\omega}(r_+-r_-)}\bigg(\frac{r_-}{r_+}-
\big(\frac{r_-}{r_+}\big)^{3\omega}\bigg)\Bigg].
\end{equation}
By attention to these expressions the total growth rate (2.15)
could be rewritten as follows.
\begin{equation}
\frac{d\mathcal{S}}{dt}=-q_E^2\big(\frac{1}{r_+}-\frac{1}{r_-}\big)+\frac{a}{2}\big(\frac{1}{r_+^{3\omega}}-\frac{1}{r_-^{3\omega}}\big)$$$$
-\frac{1}{12\omega}\bigg(\lambda+\frac{\mu}{3\omega}\bigg)\bigg(\frac{1}{r_+^{3\omega}}-\frac{1}{r_-^{3\omega}}\bigg)-
\frac{\mu}{12\omega}\bigg(\frac{\ln(r_+)}{r_+^{3\omega}}-\frac{\ln(r_-)}{r_-^{3\omega}}\bigg).
\end{equation}
Looking to the works presented by Brown et al [2,3], one can infer
 there are some extra terms due to
the presence of quintessence dark energy.
 If we rewrite this expression with respect to thermodynamic variables we find
\begin{equation}
\frac{d\mathcal{S}}{dt}=(M-\mu_+q_E-\mathcal{A}_+a)-(M-\mu_-q_E-\mathcal{A}_-a)$$$$
-\frac{1}{12\omega}\bigg(\lambda+\frac{\mu}{3\omega}\bigg)\bigg(\frac{1}{r_+^{3\omega}}-\frac{1}{r_-^{3\omega}}\bigg)-
\frac{\mu}{12\omega}\bigg(\frac{\ln(r_+)}{r_+^{3\omega}}-\frac{\ln(r_-)}{r_-^{3\omega}}\bigg),
\end{equation}
in which $\mu_\pm=q_E/r_\pm$ stands for chemical potential,
$\mathcal{A}_\pm=-1/2r_{\pm}^{3\omega}$ is conjugated potential
for parameter ``$a$". As we expect for $a=0$ the second line in
the above result vanishes. If we take ``$E$" for the average
energy of the quantum states then the rate of quantum complexity
satisfies the Lloyd bound [7] as
\begin{equation}
\frac{d\mathcal{C}}{dt}\leq\frac{2E}{\pi\hbar}.
\end{equation}
This satisfaction arises from the conditions $\mu>0, \lambda>0$
and $\lambda+\frac{\mu}{3\omega}>0$ which  are mentioned in the
above for the quintessence regime $-1<\omega<-\frac{1}{3}.$

\section{Butterfly effect  with shock wave geometry}
The shock wave geometry happens when our black hole solution is
perturbed by a small amount of energy. Study of the shock wave
geometry can be done by calculating the butterfly velocity which
is the velocity of shock wave near the horizon. To do so we first
rewrite the black hole solution (2.5) in the Kruskal coordinates
system such that
\begin{equation}
ds^2=-2H(u,v)dudv+h(u,v){d\vec{x}_2}^2,
\end{equation}
where we defined
\begin{equation}
H(u,v)=-\frac{4}{uv}\frac{f(r)}{[f^{\prime}(r_h)]^2},
\end{equation}
and $h(u,v)=r^2$  where and from now on we mark outer horizon by
$r_h$ instead of $r_+$ for simplicity reasons. As we know that
there are the following relationship between the null Kruskal
coordinates and the spherical coordinates.
\begin{equation}
u=e^{\frac{2\pi}{\beta}(-t+r_*(r))},~~~\text{and}~~~v=-e^{\frac{2\pi}{\beta}(t+r_*(r))},
\end{equation}
with the thermodynamic parameter $\beta=1/k_BT$ in which $T$ is
temperature and $k_B$ is the Boltzmann constant. Also $r_*=\int
\frac{dr}{f(r)}$ called as the tortoise spatial radial coordinate.
For neighborhood of the exterior horizon $r_h$ the tortoise
coordinate is approximated with the following form.
\begin{equation}
r_*\approx\frac{1}{f^{\prime}(r_h)}ln\big(\frac{r-r_h}{r_h}\big)+\cdots.
\end{equation}
Now by rewriting the metric (3.1) in the new coordinates we can
study the effects of disturbance as a shock wave geometry.
Actually when a scalar operator $W$ acts on the boundary at
$t_w<0$ this shock wave creates. If $t_w$ be large enough then
this shock creates a particle of null matter  which travels along
$u=0$ in the bulk. Suppose that the metric has form like (3.1) for
$u<0$ but it is changed to a perturbed metric in which $v$ is
replaced by $v+\alpha(x^i)$ [8]. $\alpha(x^i)$ is called the (red)
shift function which shows a boundary perturbation in the
direction of $x^i$.  This shift function creates some similarities
for WDW patch with unperturbed geometry at late time approximation
which is studied in previous section. In fact  when the shift
function takes some large values then the light rays of WDW patch
run into the past singularity which is similar to early time
approximation, since for small shift function similar to the late
time approximation, these light rays meet each other behind the
past horizon [3]. Applying some new transformations as
\begin{equation}
U=u,~~V=v+\theta(u)\alpha(x^i),~~\text{and}~~X^i=x^i,
\end{equation}
in which $\theta(u)$ represents the Heaviside step function, the
metric line element (3.1) then takes the new form as follows.
\begin{equation}
ds^2=-2H(U,V)dUdV+h(U,V){d\vec{x}_2}^2+2H(U,V)\alpha(X^i)\delta(U)dU^2,
\end{equation}
where $\delta(U)$ denotes to the well known ``Dirac" delta
function. It is simple to see for $u=U<0$
 the above metric reduces to old one (3.1).\\
The injected null matter stress-energy tensor, $T_{matter}$, can
be written before the injection of disturbance into the boundary
as (3.1) which in the Kruskal coordinates become
\begin{equation}
\frac{1}{\kappa}\mathcal{G}_{\{u,v\}}=T_{matter}=2T_{uv}dudv+T_{uu}du^2+T_{vv}dv^2+T_{ij}{d\vec{x}_2}^2,
\end{equation}
where $\mathcal{G}$ is the Einstein tensor. After injection this
tensor could be expressed in the new coordinates such that
\begin{gather}
\nonumber \frac{1}{\kappa}\mathcal{G}_{\{U,V\}}=T_{matter}=2(T_{UV}-T_{VV}\alpha(X^i)\delta(U))dUdV+T_{VV}dV^2\\
+(T_{UU}+T_{VV}\alpha^2(X^i)\delta^2(U)-2T_{UV}\alpha(X^i)\delta(U))dU^2+T_{ij}{d\vec{x}_2}^2.
\end{gather}
By attention to [11,41] one can consider a massless particle at
$u=0$ which moves in the $v$-direction with the speed of light,
the stress-energy
 tensor of this particle which is corresponds to the shock wave stress-energy tensor is:
\begin{equation}
T_{(shock)UU}=\frac{E}{L^4}e^{\frac{2\pi}{\beta}t}\mathfrak{a}(X)\delta(U),
\end{equation}
where $E$ is a dimensionless constant and $\mathfrak{a}(X)$ is a
local source of perturbation which for simplicity reasons we take
to be as Dirac delta function, i.e. $\mathfrak{a}(X)=\delta(X)$.
By considering the stress-energy tensor of this disturbance the
Einstein equation reads
$\frac{1}{\kappa}\mathcal{G}=T_{matter}+T_{shock}$ which should be
solved. This equation at the leading order term near the horizon
can be solved as follows.
\begin{equation}
\alpha(t,x_i)\sim e^{-\xi(|x_i|-v_Bt)},
\end{equation}
where,
\begin{equation}
\xi=\sqrt{\frac{f^{\prime}(r_h)h^{\prime}(r_h)}{2}},
\end{equation}
and $v_B$ given by
\begin{equation}
v_B=\frac{2\pi}{\beta\xi}=\sqrt{\frac{f^{\prime}(r_h)}{2h^{\prime}(r_h)}}
\end{equation}
is called as ``$\textit{butterfly~velocity}$". In fact, this
velocity as it is mentioned before is the spread of the local
perturbation on the boundary of space-time. In our case $h(r)=r^2$
and so the butterfly velocity (3.12) reads
\begin{equation}
v_B=\sqrt{\frac{\pi T}{r_h}}.
\end{equation}
Regarding the quintessence dark energy counterpart in the present
work we see that the butterfly velocity is depend on the
quintessence parameters such as normalization factor $a$ and the
state parameter $\omega$. It could be calculated by attention to
Hawking temperature $T=f^{\prime}(r_h)/4\pi$ as
\begin{equation}
v_B=\frac{1}{2}\sqrt{\frac{1}{r_h^2}+\frac{3}{L^2}-\frac{q_E^2}{r_h^4}+\frac{3a\omega}{r_h^{3\omega+3}}}
\end{equation}
in which $r_h$ is the outer horizon. To study the action growth in
this perturbed geometry two parts must be included: the action of
WDW patch behind the (I) past and (II) future horizons. By
attention to [3,22] these two parts are defined respectively by
\begin{equation}
\mathcal{S}_{future}=\frac{2M}{L\lambda_L}\int\ln
e^{\lambda_L(|t_w|-t_*+t_L-\frac{|x|}{v_B})}dx,$$$$
\mathcal{S}_{past}=\frac{2M}{L\lambda_L}\int\ln
e^{\lambda_L(|t_w|-t_*-t_R-\frac{|x|}{v_B})}dx,
\end{equation}
at which $\lambda_L$ is the Lyapunov exponent proportional to the Hawking temperature and $L$ is the length of the transverse direction $x$.
The upper bound of this coordinate called maximal transverse
coordinate is $|x|=v_B(|t_w|-t_*-t_R)$ that guarantees the emergence of shock wave effect. Time dependence of the action of WDW patch yields:
\begin{equation}
\mathcal{S}_{WDW}=\mathcal{S}_{future}+\mathcal{S}_{past}=2M(t_L+t_R)+2\mathcal{A}Mv_B(|t_w|-t_*-t_R)^2,
\end{equation}
where $\mathcal{A}$ is amplitude of shock wave in (3.10). As we
can see by disturbing the geometry the perturbation spreads on the
horizon and the action growth get corrected by an extra term which
has linear dependence to the speed of perturbation. As the shock
wave initially starts from the left side boundary of our two sided
black hole and reaches the right side so the extra part depends
only on $t_R$. By vanishing any perturbation term
$\mathcal{A}\rightarrow0$ one can re-derive non-perturbative
situation which has same rate of growth with respect to both $t_L$
and $t_R$.

Now it would be useful to study the effect of dark energy on the
butterfly velocity for the same gravity model. As we can see dark
energy leads to an extra term to $v_B$ which is addressed as the
last term in (3.14). Since $-1<\omega<-1/3$ so this term has
negative sign. Horizon radius $r_h$ as we know is a solution of
$f(r_h)=0$. In a charged black hole solution with no dark energy
around it we should set $a=0$ in the equation (2.6) as
\begin{equation}
\tilde{f}(r)=1-\frac{2\tilde{M}}{r}-\frac{\Lambda r^2}{3}+\frac{\tilde{q_E}^2}{r^2}.
\end{equation}
and so the corresponding butterfly velocity $\tilde{v}_B$ will be
obtain from (3.14) without the last term and with different
horizon radius $\tilde{r}_h$ obtained from
$\tilde{f}(\tilde{r}_h)=0.$ From (2.6) and (3.17) it is simple to
conclude that for fixed mass $M=\tilde{M}$ and fixed charge
$q_E=\tilde{q}_E$ we have $f(r)<\tilde{f}(r)$ because
$f(r)=\tilde{f}(r)-\frac{a}{r^{3\omega+1}}$. This equation is
situated properly for any horizon radius such as the horizon of
quintessence solution, namely $f(r_h)<\tilde{f}(r_h)$. Since
$f(r_h)=0$ then $\tilde{f}(r_h)>0$ and
 because $\tilde{f}(\tilde{r}_h)=0$ so it leads to $\tilde{f}(r_h)>\tilde{f}(\tilde{r}_h)$. This means the horizon radius of charged black hole
 solution must be greater when it is surrounded by the dark energy, $r_h>\tilde{r}_h$.
 The latter statement can be checked easily as follows:
 To do so we must be expand Taylor series expansion of the positive
 function $\tilde{f}(r_h)>0$ about the horizon radius in absence of the quintessence
 dark energy $\tilde{r}_h$ which leading order term is obtained as \begin{equation}
 \tilde{f}(r_h)\approx4\pi \tilde{T}(r_h-\tilde{r}_h)+O(2)\end{equation}
 in which we used the horizon equation of the black hole in absence of the quintessence $\tilde{f}(\tilde{r}_h)=0$
 and corresponding the Hawking radiation temperature $4\pi \tilde{T}=\tilde{f}^{\prime}(\tilde{r}_h).
 $
 Regarding positivity condition on $\tilde{f}(r),$ $\tilde{T}$ and $\tilde{f}(r_h)$ then the equation (3.18)
 satisfy the statement $r_h>\tilde{r}_h$. Substituting $a=0$ and assuming
 $\tilde{q}_E=q_E$ the equation (3.14) leads to the following form:
 \begin{equation}4\tilde{v}_B^2=\frac{1}{\tilde{r}_h^2}+\frac{3}{L^2}-\frac{q_E^2}{\tilde{r}_h^4}\end{equation} for which
 $\tilde{r}_h\geq \tilde{r}_h^{(0)}$ with \begin{equation}\tilde{r}_h^{(0)}=\frac{\sqrt{1+\frac{12q_E^2}{L^2}}-1}{\frac{6}{L^2}}.\end{equation}
 For small values of charge $\frac{q_E}{L}<<1$ the above minimal horizon radius reaches to the following limit.
 \begin{equation}\tilde{r}_h^{(0)}\to |q_E|.\end{equation} In this limit one can write the butterfly velocity (3.14)
 as follows. \begin{equation}4v_B^2\approx4\tilde{v}_B^2+\frac{3a\omega}{|q_E|^{3\omega+3}}\end{equation} where
 \begin{equation}4\tilde{v}_B^2=\frac{3}{L^2}.
 \end{equation} By attention to the conditions $a>0$ and $-1<\omega<-\frac{1}{3}$ one can compare (3.22) and (3.23) to infer $v_B<\tilde{v}_B.$
It means that the complexity action spreads on the AdS black hole
horizon with slower (faster) butterfly velocity in presence
(absence)
 of the quintessence dark energy. In the other side when $\omega$ is decreased and get closer to smallest value $\omega=-1,$
 then  the gap of the butterfly velocity
 arisen by dark energy get more decreased and for which we have $4v_B^2\to \frac{3}{L^2}-3a$ with $a<\frac{1}{L^2}
  $. In short one can infer that by decreasing value of
  the quintessence state equation parameter the butterfly velocity
and so the complexity decreased.

\section{Concluding remarks}
We studied the complexity growth rate by using ``CA" conjecture
[2,3] for a simple model of gravity when its AdS black hole
solution is surrounded by quintessence dark energy [30]. The
effects of this kind of dark energy is investigated earlier in
various works [29-34] and it seems challenging
 to see how it affects the holographic characteristics. We found some extra terms related to the quintessence dark energy are added to the total
 action growth.  Also it is proved that by attention to the conjugated potential for the quintessence parameter the Lloyd bound [7] is satisfied
 for all parameter states defined in regime of the quintessence dark energy.

We also investigate the action growth of this model for shock wave
geometry [9]. Actually when the boundary is perturbed by a small
amount of energy,
 the geometry in the bulk is affected. The local shock wave spreads near the horizon with the ``butterfly
 velocity" which could be obtained by the
 equation of motion for the new stress-energy tensor.
 In fact its form is same as of the old stress tensor but  with an extra term which comes from the shock
  wave and has only $UU$ component. It is due to a massless particle moving at null hypersurface $u=0$ with the speed of light.
 We showed that the effect of the quintessence dark energy causes to spread the shock wave
  with slower butterfly velocity near the horizon, so the complexity growth would be slower as well.

  \vskip .5cm
 \noindent
  {\textbf {References}}
\begin{description}
\item[1.] J. Maldacena, ``The large N limit of superconformal field theories and supergravity", Adv. Theor. Math. Phys.2, 231, (1998).
\item[2.] A. R. Brown, D. A. Roberts, L. Susskind, B. Swingle and Y. Zhao, ``Holographic Complexity Equals Bulk Action$?$", Phys. Rev. Lett. 116, 191301 (2016).
\item[3.] A. R. Brown, D. A. Roberts, L. Susskind, B. Swingle and Y. Zhao, ``Complexity, action, and black holes", Phys. Rev. D 93, 086006 (2016).
\item[4.] D. Stanford and L. Susskind, ``Complexity and Shock Wave Geometries" Phys. Rev. D 90, 126007 (2014),  hep-th/1406.2678.
\item[5.] L. Susskind, ``Addendum to computational complexity and black hole horizons", Fortschr. Phys.64, 44 (2016).
\item[6.] L. Susskind, ``The typical-state paradox: Diagnosing horizons with complexity", Fortschr. Phys.64, 84 (2016).
\item[7.] S. Lloyd, ``Ultimate physical limits to computation", Nature 406, 1047 (2000).
\item[8.] R. G. Cai, S.M. Ruan, S.J. Wang, R.Q. Yang, and R.H.Peng, ``Action growth for AdS black holes", JHEP 1609, 161 (2016).
\item[9.] S. H. Shenker and D. Stanford, ``Black holes and the butterfly effect", JHEP 1403, 067(2014), hep-th/1306.0622.
\item[10.] S. H. Shenker and D. Stanford, ``Multiple Shocks", JHEP 1412, 046 (2014), hep-th/1312.3296.
\item[11.] D. A. Roberts, D. Stanford and L. Susskind, ``Localized shocks", JHEP 1503, 051 (2015), hep-th/1409.8180.
\item[12.] W. J. Pan and Y. C. Huang, ``Holographic complexity and action growth in massive gravities",
 Phys. Rev. D 95, 12, 126013 (2017), hep-th/1612.03627.
\item[13.] R. G. Cai, M. Sasaki, and S. J. Wang, ``Action growth of charged black holes with a single horizon", Phys. Rev. D95, 124002 (2017).
\item[14.] M. Alishahiha, A. F. Astaneh, A. Naseh, and M. H.Vahidinia, ``On complexity for f(R) and critical gravity", JHEP 05, 009, (2017).
\item[15.] W. D. Guo, S. W. Wei, Y. Y. Li, and Y. X. Liu, ``Complexity growth rates for AdS black holes in massive gravity and f(R) gravity",Eur.Phys.J. C77, 904,
 (2017), gr-qc/1703.10468
\item[16.] Y. S. An and R. H. Peng, ``The effect of Dilaton on the holographic complexity growth", Phys. Rev. D 97, 066022 (2018), hep-th/1801.03638.
\item[17.] A. Ovgun, K. Jusufi, ``Complexity growth rates for AdS black holes with dyonic/ nonlinear charge/ stringy hair/ topological defects",
gr-qc/1801.09615.
\item[18.] E. Yaraie, H. Ghaffarnejad and M. Farsam, ``Complexity growth and shock wave geometry in AdS-Maxwell-power-Yang–Mills theory"
 Eur. Phys. J. C78, 967, (2018).
\item[19.] S. H. Hendi and B. Bahrami Asl, ``Complexity of the Einstein-Born-Infeld-Massive Black holes" gr-qc/1810.04792.
\item[20.] E. Perlmutter, ``Bounding the space of holographic CFTs with chaos", JHEP 10, 069, (2016).
\item[21.] M. Alishahiha, A. Davody, A. Naseh, and S. F. Taghavi, ``On butterfly effect in higher derivative gravities",
JHEP 11, 032, (2016).
\item[22.] Y. G. Miao and L. Zhao, ``Complexity/Action duality of the shock wave geometry in a massive gravity
theory", Phys. Rev. D 97, 024035 (2018), hep-th/1708.01779.
\item[23.] S. A. Hosseini Mansoori and M. M. Qaemmaqami, ``Complexity Growth, Butterfly Velocity and Black hole Thermodynamics",
 hep-th/1711.09749.
\item[24.] S. Chen, Q. Pan and J.
Jing, ``Holographic superconductors in quintessence AdS black
 hole spacetime", Class. Quant. Grav. 30, 14, 145001 (2013).
\item[25.] X. M. Kuang and J. P. Wu, ``Effect of quintessence
on holographic fermionic spectrum",  Eur. Phys. J. C 77, 670
(2017).
\item[26.] LI, Gu-Qiang, ``Effects of dark energy on P–V criticality of charged AdS black holes", Phys.
Lett. B735, 256, (2014).

\item[27.] Shinji Tsujikawa, ``Quintessence: A Review", Class. Quant. Grav. 30, 214003 (2013); gr-qc/1304.1961.

\item[28.] L. H. Ford, ``Cosmological-constant damping by unstable scalar fields", Phys. Rev. D 35, 2339(1987).

\item[29.] Y. Fujii, ``Origin of the gravitational constant and particle masses in a scale invariant scalar-tensor theory",
 Phys. Rev. D 26, 2580 (1982).

\item[30.] V. V. Kiselev, ``Quintessence and black holes", Class. Quant. Grav. 20, 1187, (2003).
\item[31.] S. Chen, B. Wang and R. Su , ``Hawking radiation in a d-dimensional static spherically-symmetric black Hole surrounded by
quintessence", Phys. Rev. D77, 124011 (2008).
\item[32.] Y. H. Wei and Z. H. Chu, ``Thermodynamic properties of a Reissner-Nordstroem quintessence black hole",
Chin. Phys. Lett. 28, 100403 (2011).
\item[33.] S. Fernando, ``Schwarzschild black hole surrounded by quintessence: Null geodesics", Gen. Rel. Grav. 44, 1857, (2012).
\item[34.] B. B. Thomas and M. Saleh, ``Thermodynamics and phase transition of the Reissner-Nordstroem black hole surrounded by
quintessence", Gen. Rel. Grav. 44, 2181, (2012).
\item[35.] H. Ghaffarnejad, E. Yaraie and M. Farsam, ``Quintessence Reissner Nordstrom anti de Sitter black holes and
Joule Thomson effect", Int. J. Theor. Phys. 57, 1671, (2018).

\item[36.] M. Olivier and T. Harko, ``New derivation of the Lagrangian of a perfect fluid with a
barotropic equation of state", Phys. Rev. D86, 087502 (2012).

\item[37.] L. Lehner, R. C. Myers, E. Poisson and R. D. Sorkin, ``Gravitational action with null boundaries", Phys. Rev. D94 ,
084046 (2016).
\item[38.] D. Carmi, S. Chapman, H. Marrochio, R. C. Myers and S. Sugishita,
``On the time dependence of holographic complexity", JHEP 2017,
188 (2017).
\item[39.] Y. R. Qiu, C. Niu, C. Y. Zhang and K. Y. Kim, ``Comparison of holographic and field theoretic complexities for
time dependent thermofield double states", JHEP 2018, 82, (2018).

\item[40.] Gao, Changjun, Youjun Lu, Shuang Yu, and You-Gen Shen.
``Black hole and cosmos with multiple horizons and multiple
singularities in vector-tensor theories", Phys. Rev. D97, 104013
(2018).

\item[41.] K. Sfetsos, ``Gravitational shock waves in curved space-times", Nucl. Phys. B436, 721 (1995), hep-th/9408169.

\end{description}
\end{document}